\documentclass[twocolumn]{aa}

\usepackage{graphicx} 
\usepackage{subcaption}
\usepackage[colorlinks=true, linkcolor=teal, citecolor=teal, urlcolor=gray]{hyperref}
\usepackage{xcolor}
\usepackage{listings}
\usepackage{amsmath}
\usepackage{xspace}
\usepackage{changes}

\bibpunct{(}{)}{;}{a}{}{,}

\begin{document}
 
\title{Visual analytics for cosmological simulation results}
\author{Paul Vauterin and Maarten Baes}
\date{July 2026}

\institute{Department of Physics and Astronomy, Universiteit Gent, Proeftuinstraat 86 N3, B-9000 Gent, Belgium. \\ \email{paul.vauterin@gmail.com}}

\abstract{
\textbf{Context:} Modern cosmological simulations rely on sophisticated assumptions and generate rich, highly complex datasets. Interpreting their results in order to extract new scientific insights remains challenging. Existing visualisation tools offer powerful capabilities but often come with a relatively high barrier for entry, and are not designed with a focus on knowledge discovery through real-time, intuitive visual analytics. This stands in contrast to other research domains, such as bioinformatics, where visual analytics tools have become deeply embedded in scientific discovery workflows.

\textbf{Aim:} Our goal is to complement the existing ecosystem of cosmological visualisation tools with a lightweight, user-friendly application that supports visual analytics and frictionless dissemination of results to the scientific community.

\textbf{Methods:} We developed ARGOS, an open-source, web-based environment for real-time visual analytics, tailored to cosmological simulation outputs and designed with an emphasis on user experience. ARGOS combines GPU-accelerated browser rendering for interactive exploration of large datasets with a template-driven approach that enables rapid adaptation to other types of data and analysis workflows. Importantly, ARGOS combines catalogue-level and object-level exploration, allowing users to move seamlessly from large ensembles of objects to snapshots of individual objects.

\textbf{Results:} We demonstrate how ARGOS can support intuitive visual data exploration through example dashboards for multiple cosmological particle datasets, and illustrate broader applicability with dashboards for SKIRT synthetic multi-band imaging data products. Source code is freely available at \url{https://github.com/pvaut/skirt-argos} under the MIT licence. A reference deployment, including sample data, is available at \url{https://skirt-argos.ugent.be}.
}


\maketitle
\nolinenumbers

\section{Introduction}

Understanding the formation and evolution of cosmic structures increasingly relies on large-scale, sophisticated numerical simulations, modelling physical effects that influence the evolution of the universe, such as gravity, hydrodynamics, and various feedback mechanisms
(for reviews, see \citet{SomerVilleReview}, \citet{VogelsbergerReview}, \citet{Crain_2023}, \citet{Feldmann2025CosmologicalSimulations}). Such simulations now routinely generate billions of particles over hundreds of time steps, capturing the detailed distribution of dark matter, gas, and stellar components. These results have strongly shaped our theoretical picture of structure formation. However, their ever-increasing sophistication has also created a growing challenge: the volume, complexity, and diversity of the outputs make it difficult for scientists to inspect results, diagnose issues, and develop intuition about the numerical modelling of underlying physical processes.

Evidently, data visualisation is a key tool to address this challenge. Existing visualisation tools, including Splotch \citep{dolag2008splotch}, yt \citep{turk2011yt}, TOPCAT \citep{topcat2017}, Topsy, ParaView and VisIt provide powerful capabilities and are widely used. 
What these tools do not yet offer is an approach that brings visual analytics, as exemplified by linked-view systems such as Glue \citep{beaumont2015glue}, to the exploration of cosmological simulation data.

This gap is increasingly limiting. As simulations get ever more sophisticated, researchers face more complex potential failure modes and subtle emergent structures that may not be discovered through predetermined analysis flows and plots alone. Several scientific fields confronted with complex datasets have adopted real-time, intuitive visual analytics as a transformative methodology. Astronomy itself has a long tradition of interactive data exploration tools, including TOPCAT \citep{topcat2017}, Firefly \footnote{\url{https://github.com/Caltech-IPAC/firefly}}, Vaex \citep{vaex}, and tools developed within the International Virtual Observatory Alliance (IVOA). These systems demonstrate the value of interactive visualisation for astronomical research, particularly in the exploration of observational datasets such as catalogues, images, and spectra.

Beyond astronomy, visual analytics has become central to areas of life sciences that produce complex data such as epidemiology \citep{Chui2011_VAEpi,Vauterin2017_Panoptes}, single-cell transcriptomics (e.g., Cytosplore Viewer \footnote{\url{https://viewer.cytosplore.org}}, \citet{hollt2016cytosplore}) and spatial multi-omics \citep{zhou2021omicsanalyst, keller2025vitessce}. These bioinformatics platforms provide user-friendly dashboards with multiple linked views that reveal complementary aspects of the same dataset and allow interactive visual filtering, enabling researchers to detect patterns and anomalies in complex, high-dimensional data and to assess data quality and generate hypotheses. Another example is social sciences, where visual analytics tools are widely used to explore large volumes of geospatial data and networks (e.g. \citet{Pezanowski2018_SensePlace3}, \citet{VAsocialnetworks}). 

Cosmological simulations, despite facing analogous challenges, lack a comparable interactive visual analytics environment tailored to their specific requirements, such as the interactive selection of regions in space and the joint exploration of three-dimensional spatial and velocity structure in phase space.

To address this need, we present ARGOS, a browser-based software application for intuitive, interactive visual analytics of cosmological particle simulations. ARGOS runs entirely on the client side using WebGL, enabling high-performance rendering and interaction without installation, compilation, or server-side computation. By eliminating platform friction and emphasising an engaging user experience, ARGOS makes visual analytics immediately accessible to a broad community of researchers and students, lowering barriers to exploratory analysis and accelerating discovery of insights.

ARGOS was designed around the following core objectives:

\begin{itemize}
\item
User-friendly: intuitive to learn, responsive, and enjoyable to use.
\item
Flexible: adaptable to diverse datasets and workflows, and extensible for project-specific needs.
\item
Powerful: capable of rendering and interacting with millions of data points in real time, including strong visual analytics features.
\item
Easy to deploy: platform independent, small footprint, and very low maintenance.
\end{itemize}

ARGOS' mission statement is real-time, user-friendly browser-native visual exploration. This focus brings clear trade-offs: ARGOS does not aim to replace HPC-native or desktop tools for heavy 3D volume rendering, in-situ analysis, or server-side querying of arbitrarily large datasets. Real-time rendering performance in the browser ultimately depends on client hardware and browser memory limits, and workflows that require complex computations over full-resolution data necessitate external preprocessing. We therefore position ARGOS as a complementary tool optimised for rapid, low-barrier and user-friendly exploratory analysis and dissemination of data products that are preprocessed to match these limitations for interactive work. In case data sets are too large for ARGOS, an usually acceptable workflow is to downsample the data prior to upload, a standard operation that could be done by a simple Python script using the \texttt{h5py} package. 

Because flexibility was a central objective, ARGOS can also be used to explore data beyond cosmological simulations. In practice, it applies to any dataset representable as one or more tables. One example is multi-band imaging, which can be structured as a table with rows corresponding to pixels and columns representing wavelength bands. We demonstrate this broader applicability by applying visual analytics principles to synthetic images generated with the SKIRT radiative transfer software. ARGOS also supports a powerful catalogue drill-down workflow, in which users begin by visually exploring a catalogue of objects (e.g., galaxies) and then identify specific objects of interest for which the data is downloaded on the fly and a detailed dashboard is constructed (e.g., analysing multi-wavelength synthetic images of an individual galaxy).

\section{Methods}

This section consists of two parts: subsections 2.1, 2.2 and 2.3 introduce some technical terminology and describe the software engineering foundations, whereas sections 2.4 and 2.5 focus on how ARGOS can be used.

\begin{figure*}[t]
    \centering
    \includegraphics[width=\textwidth]{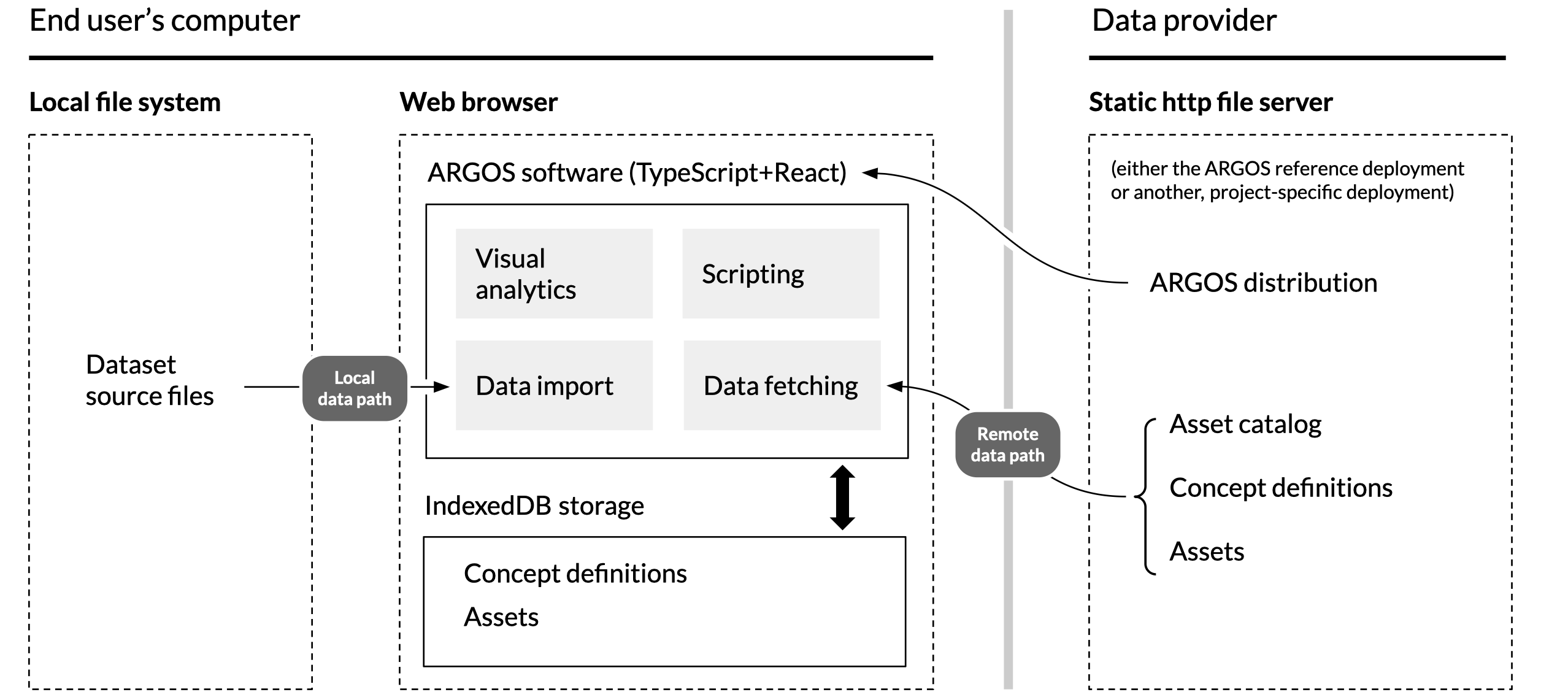}
    \caption{Schematic overview of the ARGOS software architecture and data flow. See Section~\ref{software_architecture} for details.}
    \label{fig:software_architecture_overview}
\end{figure*}

\subsection{Data architecture}

To maximise flexibility, ARGOS is designed to be largely agnostic about both the type of data it visualises and the way that data is shown. Its data architecture revolves around five core elements: concepts, assets, tables, properties, and entities.

A concept defines a class of datasets (assets) that share a common structure, meaning they contain the same tables and properties. A concept specifies:
\begin{itemize}
    \item how a dataset is processed during import to create an asset,
    \item which derived properties should be computed,
    \item how the data should be visualised in the associated dashboard, and
    \item optional scripts that automate user-triggered actions.
\end{itemize}
Much of ARGOS's flexibility comes from the ability to configure concept definitions for new data types. Technically, each concept is stored as a YAML file and can be edited directly. In addition, the front end provides user-friendly editors for all parts of a concept definition, including, for example, a visual dashboard builder.

An asset is ARGOS's internal representation of a single dataset, displayed in one dashboard. An asset contains one or more tables. Typically, data are imported into an asset from an HDF5 file. For instance, an asset might represent one snapshot of a cosmological simulation, with separate tables for gas, stellar, and dark matter particles.

A table consists of rows, called entities that correspond to individual data points, and columns, called properties.

\subsection{Software architecture} \label{software_architecture}

Figure~\ref{fig:software_architecture_overview} provides a schematic overview of the ARGOS architecture and data flow. At its core, ARGOS is a client-only web application written in TypeScript and built with the React framework. All data required to display dashboards, including concept definitions, assets, and tables, are stored in IndexedDB, a browser-embedded NoSQL database accessible to web applications (note that size limitations and clearing of this database can be managed via the browser settings). Data can be added to ARGOS through two complementary paths: a local path and a remote path.

\subsubsection{Local data}

In the local data path, users import files from the file system on their own computer. This mode is most useful for researchers generating data who wish to troubleshoot or explore their results. Because the data resides on the user’s computer, loading and interacting with it is fast, comparable to a native desktop application.

\subsubsection{Remote data}

In the remote data path, a data provider deploys ARGOS and serves datasets statically over HTTP from the same server environment. This mode is particularly useful for consortia (e.g., cosmological simulation collaborations) that want to share their results with the wider community through ARGOS. Crucially, due to ARGOS's architecture, this approach does not require running server-side code. As a result, it is well suited for long-term “deploy-and-forget” projects, since static HTTP hosting is ubiquitous, inexpensive, scalable, and stable without ongoing maintenance such as security patching.

For remote deployments, an asset catalogue, configured as a YAML file, must also be provided so that the web client can discover which concepts and assets are available. The inline documentation of ARGOS describes the required folder structure and the format of this catalogue.

Because all datasets are stored locally in IndexedDB after first load, remote data only need to be fetched once. Given that ARGOS can handle datasets exceeding 1~GB, this caching behaviour is important for usability. Whenever ARGOS starts, it checks the served asset catalogue for newly available or updated assets, using a version tag to detect changes.

The two data paths are not mutually exclusive: a researcher can work with remote datasets from a shared ARGOS deployment while also importing additional local data using the same deployment.

\subsection{Scientific software engineering patterns}

Beyond its direct use as a visualisation tool, the codebase may also serve as a source of inspiration for researchers developing other scientific web applications, as it implements several non-standard architectural patterns that are broadly applicable for scientific software.

One such pattern is the client-side storage of datasets in IndexedDB, used consistently for both locally imported data and data fetched from a static HTTP server. This design reduces repeated downloads, enables fast reloading, and keeps the deployment footprint minimal.

A second pattern is ARGOS's flexible and extensible dashboard configuration and rendering engine. The system relies on factory patterns and standardised object interfaces, allowing new plot types, interactions, and layouts to be added with limited coupling to the rest of the code.

A third pattern is the inclusion of a lightweight client-side scripting engine that supports end-user customisation and task automation. Similar mechanisms could benefit many scientific web applications by dramatically increasing their versatility without requiring users to modify source code.

Regarding high-performance data handling in the browser, ARGOS contains several implementation choices that may serve as inspiration:
\begin{itemize}
\item
Low-level, transparent use of WebGL. ARGOS interacts directly with the WebGL API rather than relying on higher-level libraries such as \url{threejs.org}. This choice maximises performance for the specific rendering tasks at hand, minimises unnecessary transfers between CPU and GPU memory, and enables the use of shaders precisely tailored to the application (e.g. for the dynamic colour coding by line-of-sight velocity).
\item
Efficient in-memory data structures. Large arrays are stored and manipulated in JavaScript using low-level \texttt{TypedArray} objects, which provide compact storage and predictable performance for numerical workloads.
\end{itemize}

\subsection{Data import}
\label{section:data_import}

\subsubsection{Supported file formats}

The primary import format in ARGOS is HDF5, which is effectively the standard for cosmological simulation data. Widely used simulation codes such as GADGET-4 \citep{Springel2021_Gadget4} and SWIFT \citep{Schaller2024_SWIFT} write their snapshot outputs in HDF5, and recent I/O frameworks describe HDF5 as the “de facto standard” for large-scale simulations \citep{Hausammann2022_CSDS}. ARGOS uses the h5wasm library \footnote{\url{https://github.com/usnistgov/h5wasm}}, a fast and scalable WebAssembly-based implementation of HDF5 I/O in JavaScript.

As an additional convenient import option for smaller and simpler datasets, ARGOS also supports TAB-delimited text files.

\subsubsection{Data augmentation}

Simulation outputs are often highly technical, which can hinder understanding and interpretation. For example, HDF5 datasets frequently use technical identifiers instead of human-readable names, and values are typically stored in computational units rather than in astrophysically meaningful units.

To reduce this barrier, ARGOS supports data augmentation during import through two mechanisms: (1) adding metadata and (2) defining transformations. These steps can be configured via a user-friendly wizard and are stored at the concept level, so that ARGOS automatically applies them to every asset imported into that concept.

For metadata, users can assign readable names to properties, add descriptions and unit labels, and group related properties into sections to improve interpretability. This metadata is then used throughout the interface to guide users and provide context.

In addition, each imported property can be associated with a transformation, specified as a generic mathematical expression. These transformations can be used to convert values from computational to astrophysical units at import time.

\subsubsection{File size limitations}

Web browsers impose practical limits on the maximum size of in-memory objects. Because ARGOS loads all data for an asset into memory, this translates into a maximum asset size of roughly 1--2~GB on current-generation browsers.

This is far smaller than the raw outputs of many cosmological simulations. However, this limitation does not conflict with ARGOS's goals: its purpose is to enable real-time, interactive visual analytics for troubleshooting and understanding simulation results. Achieving responsive rendering and real-time visual analytics interaction on a typical modern laptop requires limiting the particle count to on the order of $10^7$. Datasets of this size generally fit within the $\sim$1~GB browser memory budget.

For larger simulation outputs, a straightforward solution is to randomly subsample particles prior to import. Subsampling preserves ARGOS's utility for interactive exploration while keeping datasets within browser limits and facilitating real-time rendering. The ARGOS source repository includes a sample Python script to perform this subsampling.
 
\begin{figure*}[t]
    \centering
    \includegraphics[width=\textwidth]{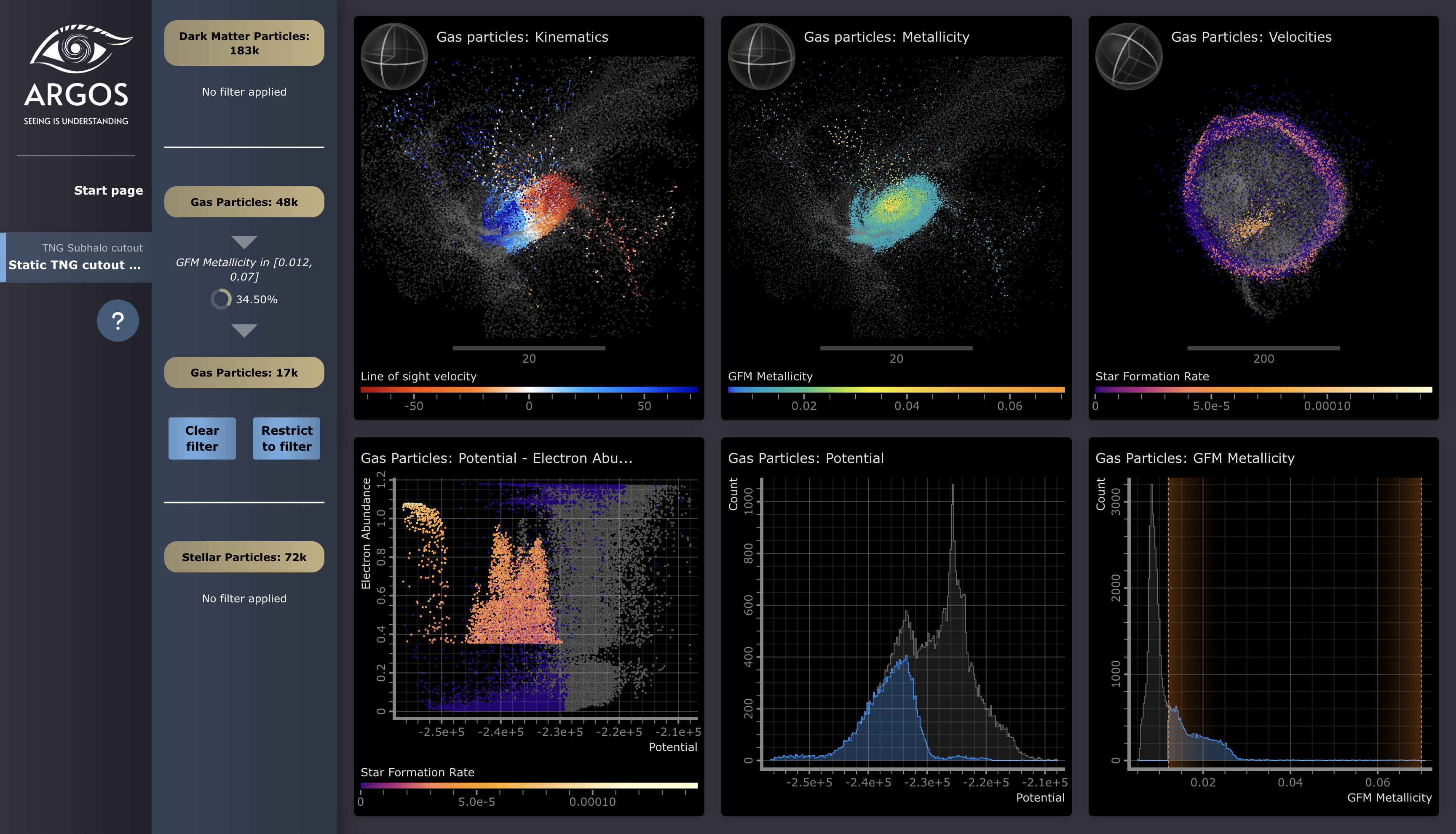}
    \caption{Example of an ARGOS dashboard showing gas particles from a TNG50 simulation cutout. Top row, left to right: spatial view with particles colour-coded by line-of-sight velocity and velocity trails indicated; the same spatial view colour-coded by metallicity; particles in velocity space colour-coded by star formation rate. Bottom row: scatterplot comparing two properties and colour-coded by a third; histogram of the potential; histogram of the metallicity. A metallicity filter was created and applied to all plots, with unselected particles shown in grey. The left sidebar lists all active filters.}
    \label{fig:argos_dashboard_1}
\end{figure*}

\subsection{Visual analytics dashboards}

A key focus of ARGOS is advanced visual analytics \citep{thomas2005illuminating,keim2010visual}: interactive techniques that support scientific reasoning rather than purely aesthetic visualisation. Building on the long tradition of coordinated multiple views (e.g., \citet{north2000snaptogether}, \citet{heer2012interactivedynamics}), the software integrates brushing and linking, dynamic filtering, and details-on-demand exploration across multiple synchronised plots that present different aspects of the same dataset.

These interactions allow users to select subsets of particles in one view, for example, a spatial plot or scatterplot, and immediately observe the corresponding distributions in other views. This facilitates the identification of correlations and anomalies in complex, high-dimensional datasets. The system therefore functions not only as a visualisation engine, but also as a tool for troubleshooting and understanding simulation outcomes, revealing artefacts or emergent structures that might otherwise remain hidden, and to generate hypotheses via data exploration. Figure~\ref{fig:argos_dashboard_1} illustrates an ARGOS dashboard with multiple coordinated views showing the same particle selection.

\subsubsection{Configurable dashboards}

The interface design of ARGOS emphasizes user-assembled dashboards. Inspired by visual composition frameworks such as Snap-Together visualisation \citep{north2000snaptogether} and Polaris \citep{stolte2002polaris}, users can arrange multiple coordinated plots, maps, and histograms into reusable dashboard templates. These templates are automatically applied to all assets that belong to the same concept. This approach supports reproducible visual analysis: a researcher can construct a custom layout tailored to a specific dataset or scientific question, export it, and share it with collaborators, who can then recreate the same visual environment directly in the browser.

\subsubsection{Interactive filtering}

ARGOS maintains a current selection of particles, defined by a filter query created through user interactions with the plots in a dashboard. Every plot automatically reflects this selection; for instance, in a scatterplot, non-selected points are greyed out. This central visual analytics feature provides a powerful way to explore large, high-dimensional datasets. Through multi-facet reasoning, users can compare relationships, identify subpopulations and outliers across dimensions, and generate hypotheses by viewing the same selection from multiple perspectives.

User actions that generate filter queries include range selection on histograms, lasso selection in scatterplots or 3D spatial plots, and clicking items in categorical facets. Subsequent filters are combined using logical AND. At any point, users can inspect the active filter steps in a sidebar. These steps can be edited and exported in a self-documenting YAML format, for re-import into ARGOS or use in other applications.

\begin{figure}[ht]
    \centering
    \includegraphics[width=\columnwidth]{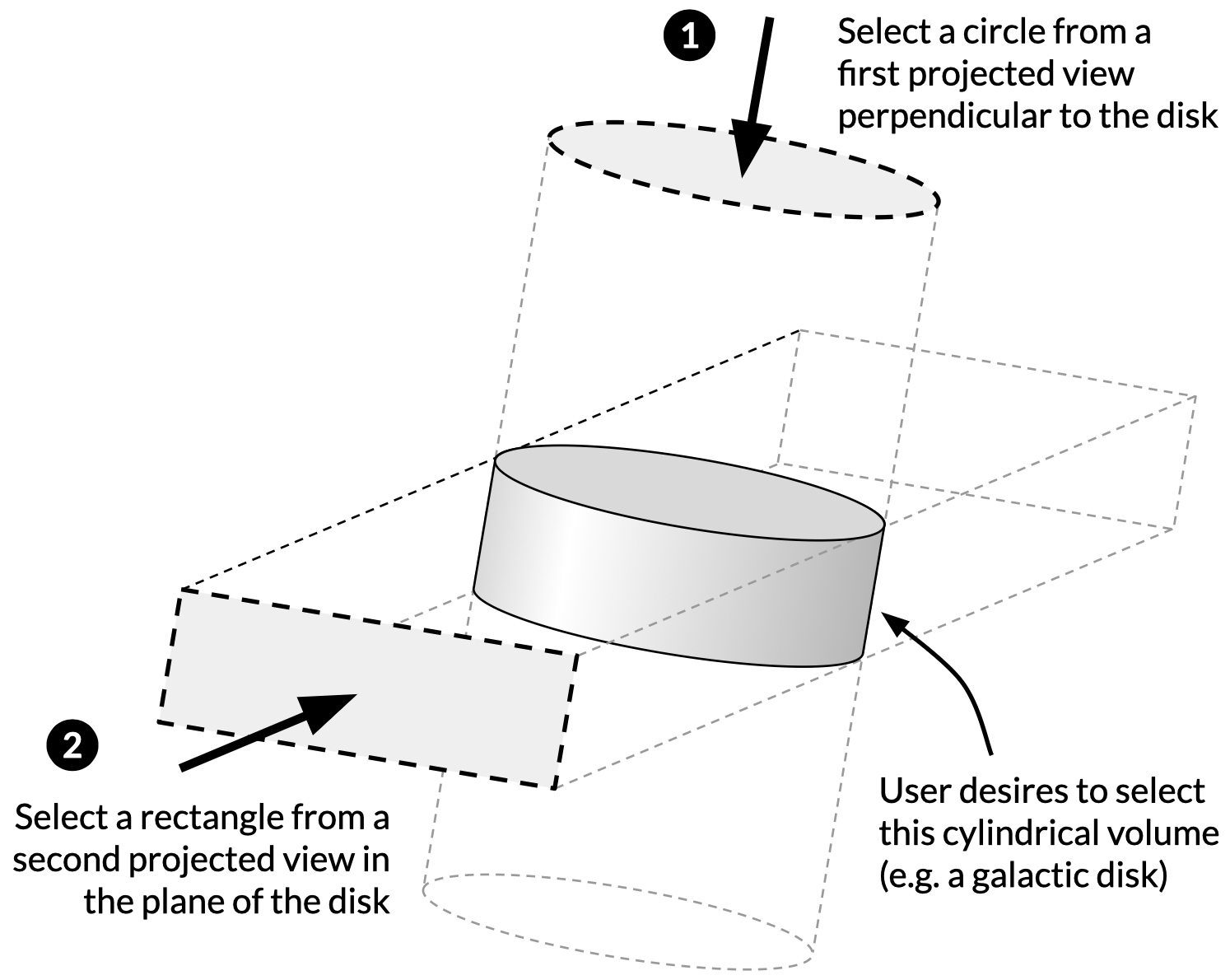}
    \caption{Selecting a specific spatial volume using two successive lasso selections from different viewing directions (here forming a cylinder). By iteratively narrowing the selection from multiple viewing angles, users can visually constrain arbitrary spatial subpopulations.}
    \label{fig:selection_3d}
\end{figure}

For cosmological simulations, a particularly important capability is the visual selection of particles within a specific region of space. For instance, a user may wish to isolate particles in a disc, spiral arm, or satellite galaxy. A single lasso selection is insufficient, because it is performed on a projected view, and therefore corresponds to a right prism of infinite length in 3D. To address this, ARGOS allows users to combine multiple lasso selections from different viewing directions. This provides a simple, intuitive, yet powerful method for interactively selecting finite 3D volumes (see Fig.~\ref{fig:selection_3d}).

\subsubsection{Connected views}

It is often useful to display more than one spatial view, for example to compare different components (dark matter, gas, stars) or to show the same component with different properties overlaid. To support this, multiple spatial plots can be synchronised so that changes to the viewpoint in one plot automatically propagate to the others.

\subsubsection{Highlighting individual data points}

It is sometimes useful to be able to inspect individual entities. If a user clicks on a point in a view (e.g. a scatterplot), this point becomes highlighted in all plots, and a sidebar pops up with all information available for that entity.

\subsubsection{Real-time visualisation}

Responsiveness, with update latencies below 500~ms, is widely regarded as crucial for interactive visual exploration, because it encourages users to iterate, probe, and explore more deeply \citep{liu2014latency,piringer2009multithreading}. In ARGOS, this level of responsiveness is achieved in the browser through a combination of technical choices.

Most importantly, visualisations are rendered on the GPU using WebGL, and the architecture minimises data transfers between the JavaScript context and GPU memory. One example is the slicing feature in ARGOS: users can specify a slicing property for a plot so that only points within a movable window of that property are shown. This is ideal for scanning through a third dimension in a 2D projection. The slicing is implemented entirely in the vertex shader, with all source data loaded into GPU buffers, avoiding any round trip to JavaScript when updating the visible subset.

In addition, code that runs on the CPU rather than the GPU (e.g., lasso selection) is implemented with careful attention to performance, sometimes departing from standard React best practices. For example, data are stored in low-level TypedArray objects instead of conventional JavaScript structures and are not maintained in the Redux store.

\subsubsection{Plot types}

A 3D kinematics plot displays a set of points using their 3D coordinates and 3D velocities. The velocities are used to colour the points according to their line-of-sight velocity from the current viewpoint. This computation happens real-time when the user changes the point of view, thanks to computation in a GPU vertex shader. The velocity is also used to show a small trail for each point to give a visual clue regarding the direction of movement of the point.

A 3D scatterplot displays a set of points using their 3D coordinates and an optional additional property that is used to colour code the points.

A 2D scatterplot displays a set of points using two numerical properties as X and Y axis, an optional additional property that is used to colour code the points.

For 3D kinematics views, 3D and 2D scatterplots the user can optionally specify an additional numerical property to define a slicing window that can be modified interactively.

 A strip plot combines a categorical property, used as bins on the X axis with a numerical property on the Y axis, an optional additional property that is used to colour code the points.

A categorical facet lists all states of a categorical property.

A histogram displays the distribution of values of a numerical property.

A parallel coordinates plot draws a number of numerical properties on parallel vertical axes, and connects the values that correspond to the same points.

\subsubsection{Customisations}

A user can easily define a derived property that is computed from the properties that are present in the source data set. Derived properties are defined as mathematical expressions and can be stored in the concept definition so that they are automatically applied to new assets belonging to that same concept. This can be used for simple transformations like $log$ conversion, but also for more complex operations, including 3D vector calculations. New derived properties can be created interactively during a data exploration session, and new plots that use them can be added on the fly.

ARGOS also contains a basic script language that can be used to add custom automated actions to a dashboard. Script can be attached to a custom button that appears either in the dashboard, or at the level of an individual entity opened in the sidebar. The script editor contains a lightweight embedded help, and a syntax validator. Sections \ref{example:skirt} describes a concrete illustration of this pattern, available in the showcase data that is part of the reference deployment.

ARGOS uses a parser library to interpret expressions and evaluate them safely and robustly. For derived properties, an efficient TypeScript implementation ensures near real-time computation, even for millions of data points.

\subsubsection{User experience}

Systems that support rapid and expressive manipulation have been shown to increase engagement and broaden the range of insights generated during exploratory tasks \citep{heer2012interactivedynamics,pike_science_of_interaction}. A central goal of ARGOS is therefore to provide an engaging user experience that encourages learning, sparks curiosity, and supports deeper exploration. Key contributing elements include:

\begin{itemize}
\item
Responsiveness. Because all data are loaded into the browser’s JavaScript memory (and often into GPU memory), interactions execute quickly and require no server round trip.
\item
Consistency. Different chart types follow a shared logic for presenting configuration options and expose common features whenever applicable (e.g., the options for colour-coding by a property).
\item
Refinement. The user interface has been polished beyond typical academic software standards, improving clarity and engagement.
\end{itemize}

\section{Results \& example use cases}

ARGOS is publicly available to the scientific community. The source code can be obtained from \url{https://github.com/pvaut/skirt-argos} and is released under the MIT licence. 

A reference deployment of ARGOS is available at \url{https://skirt-argos.ugent.be/}. Researchers can use this deployment to import and explore their own datasets via the mechanisms described in Section~\ref{section:data_import}. The deployment also includes several example datasets that demonstrate the interactive exploration features of ARGOS, and its ability to serve large scientific datasets to the community in an accessible yet powerful way. These examples are not intended as standalone scientific results, but rather as illustrations of representative usage patterns. For example, users can drag and drop any IllustrisTNG or EAGLE snapshot of interest and explore it in a manner analogous to the examples presented here.

\subsection{Cosmological simulation particle sets}

The assets \texttt{TNG282784} and \texttt{TNG479290} contain galaxy-scale cutouts from the IllustrisTNG \citep{Nelson2019TNGdata}. \texttt{EAGLE snapshot 028} contains the full simulation box of the small-scale reference simulation RefL0012N0188 from the EAGLE project \citep{Schaye2015EAGLE}. These examples illustrate the core use case of ARGOS: interactive exploration of cosmological particle data. Each asset includes particle tables for different components (gas, stellar and in some cases dark matter), along with a selection of key properties such as mass, metallicity, velocity, and star formation rate. Together, they illustrate how coordinated views and interactive filtering can be used to identify substructures, correlations, and improve overall understanding of complex simulation outputs.

The asset \texttt{TNG300 Subhalo Catalogue z=0} also derives from IllustrisTNG, but visualises the full simulation volume at redshift $z=0$, with each data point representing a subhalo rather than an individual particle. With approximately 14 million data points, this dataset highlights the scalability of ARGOS for real-time interaction with larger data sets.

\subsection{SKIRT synthetic images}
\label{example:skirt}

\begin{figure*}[t]
    \centering
    \includegraphics[width=\textwidth]{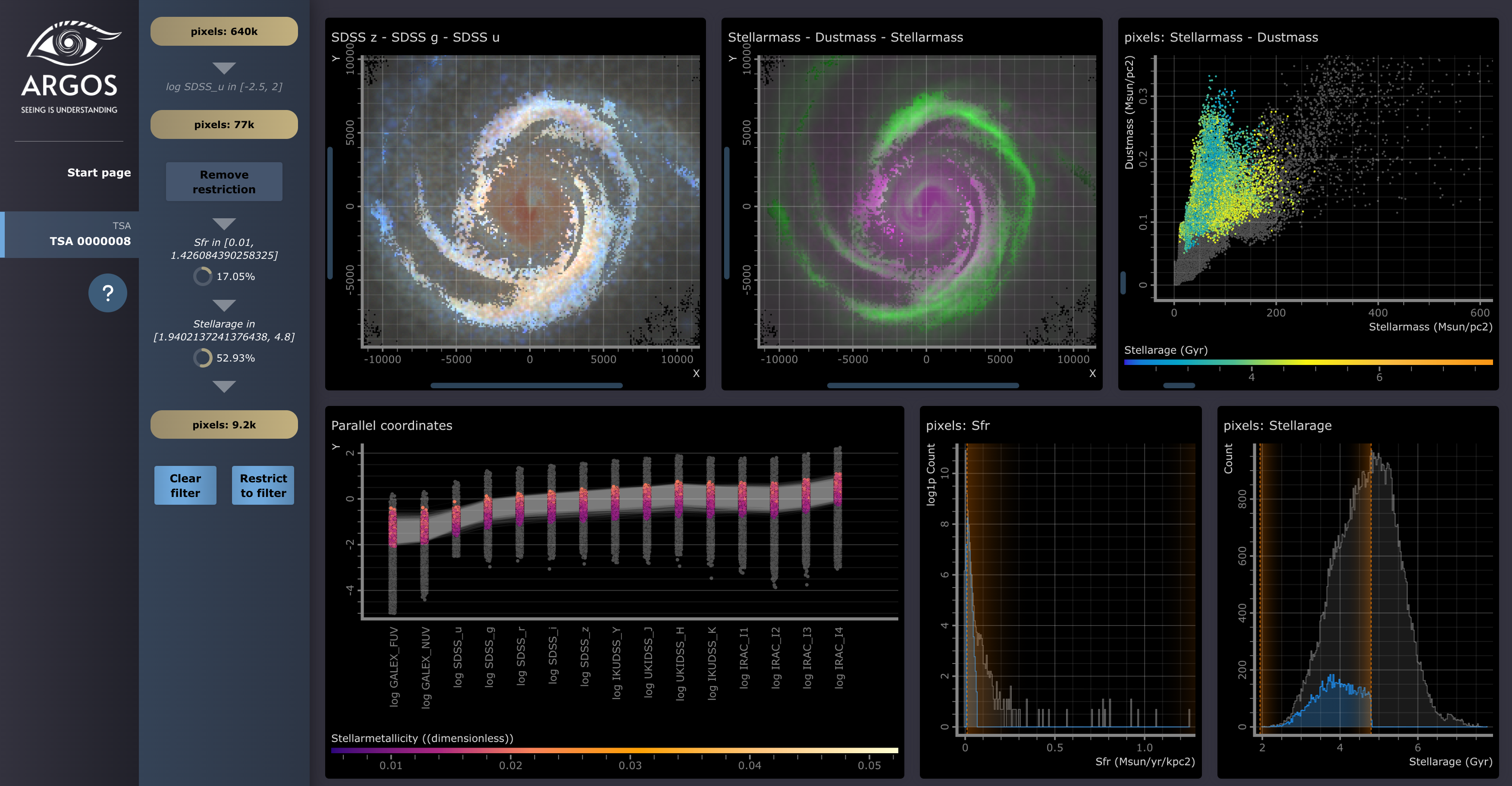}
    \caption{Example of an ARGOS dashboard showing a TNG50-SKIRT Atlas galaxy. Top row, left to right: pseudocolour image of the synthetic $u$, $g$, and $z$ bands computed via radiative transfer; pseudocolour image showing the stellar and dust mass accumulated per pixel; scatterplot of the same pixel values. Bottom row: parallel coordinates plot showing the full spectral energy distribution per pixel; histogram of star formation rate; histogram of stellar age. To remove noisy pixels in very faint regions, the dataset was first restricted based on $u$-band brightness. A subsequent filter selects pixels with ongoing star formation and an on-average older stellar population, yielding a subset that traces the spiral structure of the galaxy. On the top--left images, the unselected pixels appear in a darker, desaturated colour.}
    \label{fig:argos_dashboard_skirt}
\end{figure*}

The TNG50-SKIRT Atlas (TSA) is a synthetic, multi-wavelength image atlas for a complete, stellar-mass-selected sample of 1154 galaxies extracted from the TNG50 cosmological simulation at zero redshift. The radiative transfer post-processing was performed with SKIRT \citep{Baes2024TNG50SKIRTAtlas,Baes2024TSAWavelengthDependence}, and enables controlled comparisons between simulated and observed galaxy imaging across wavelengths.

The asset \texttt{TNG50-SKIRT Atlas} contains an overview of the full TSA galaxy catalogue. Its dashboard enables users to explore correlations among global galaxy properties and to identify objects of particular interest for follow-up analysis.

This asset provides an example of a powerful catalogue drill-down workflow that can be achieved with ARGOS:
\begin{enumerate}
    \item
    Explore a catalogue of entities. In this case, galaxies in the TSA.
    \item
    Use visual analytics (linked views, filtering, and selection) to slice \& dice the catalogue and identify individual entities of interest.
    \item
    Open a selected entity for detailed inspection as a new asset in a separate dashboard. ARGOS then dynamically fetches the corresponding data. Here, this means loading per-pixel values for the chosen galaxy across multiple wavelengths.
\end{enumerate}

Step~3 is implemented via a script defined at the concept level, which adds a custom button to the entity side panel in the atlas overview dashboard. The script fetches the associated data from the static HTTP server, and adds it to the user's local ARGOS database. By pushing this logic to a script, the ARGOS source code can be kept generic while offering flexibility for setting up such workflows.

The dashboard for an individual galaxy combines imaging data (both synthetic observables and physical, simulation-derived quantities not directly accessible to observations) with pixel-level comparison plots (see Fig.~\ref{fig:argos_dashboard_skirt}). The visual analytics features enable rich, bidirectional exploration:
\begin{itemize}
    \item
    Select a region in an image and immediately inspect how its pixels are distributed in linked scatterplots or other views.
    \item
    Select a subset of pixels in a scatterplot and highlight their spatial distribution within the image.
\end{itemize}

A concrete example of how visual analytics with ARGOS can lead to the discovery of data processing artefacts is reported in Baes et al. (in prep.). At the catalogue level, interactive visual exploration of the TNG50-SKIRT Atlas revealed a small subpopulation of galaxies that appeared as outliers in the multi-dimensional space of integrated galaxy properties. Drilling down to the corresponding synthetic image dashboards for individual galaxies exposed morphological anomalies in the dust distribution. A root-cause analysis traced these anomalies to a subtle artefact, affecting a small subset, in the dust-allocation scheme used during the SKIRT radiative transfer step. This issue has since been resolved in the second generation of the TNG50-SKIRT Atlas.

The ability to interactively explore a high-dimensional property space at the catalogue level, combined with the immediate drill-down to individual galaxies, was instrumental in identifying and resolving this rare and subtle failure mode in a complex data processing pipeline.

\section{Discussion \& conclusions}

ARGOS was developed by the research group that created and maintains SKIRT, a radiative transfer code widely used to generate synthetic observations from cosmological simulations \citep{camps_baes_2015_skirt}. Its development aligns with the group's long-term mission to provide the scientific community with powerful, robust and accessible tools that facilitate meaningful interpretation of simulations and maximise their scientific impact on cosmology and extragalactic research. By bringing real-time visual analytics to cosmological data in a low-threshold and intuitive way, ARGOS directly addresses the challenges outlined in the introduction: the growing complexity of modern simulations, the difficulty of diagnosing subtle issues, and the need for tools that support exploratory, hypothesis-generating workflows. Its applications range from inspecting and troubleshooting particle data sets to exploring synthetic images produced by the SKIRT radiative transfer pipeline.

Many excellent visualisation tools exist for cosmological particle data, some surpassing ARGOS in breadth of features and scalability. However, none of these provide a browser-native platform that combines interactive visual analytics, strong emphasis on user experience, and effortless data dissemination. ARGOS therefore offers a novel and significant contribution to the tooling ecosystem, with its distinct value lying in four main pillars.

First, ARGOS enables the construction of visual analytics dashboards composed of multiple coordinated views. Selection filters applied in one plot are immediately reflected across others, revealing relationships between different aspects of the same high-dimensional dataset. As demonstrated in fields such as life sciences and social sciences, this type of linked-view interaction is a powerful approach for uncovering trends, anomalies, emergent behaviour, or data quality issues that would remain hidden in static plots or predetermined analysis pipelines. In the context of cosmological simulations, such capabilities support model validation, help diagnose numerical artefacts, and assist researchers in building intuition about the underlying physical processes. Ultimately, this can lead to more efficient troubleshooting, fewer errors, and increased trust in simulation outcomes.

Second, ARGOS places strong emphasis on responsiveness, UI refinement, and overall user-friendliness. A defining feature is that it runs entirely in the browser, requiring no installation, compilation, or specialized hardware beyond a modern laptop or desktop. This low barrier to entry promotes adoption by the scientific community, supports reproducibility, broadens participation, and enables rapid exploratory analysis that complements more specialized HPC-based workflows.

Third, thanks to its generic and extensible design ARGOS is highly versatile and adaptable to new types of data and analysis objectives. We illustrated this flexibility through the diversity of sample datasets included in the reference deployment, ranging from particle sets to synthetic imaging data products. We expect that researchers will identify additional use cases across cosmological and non-cosmological contexts, particularly wherever tabular and spatially structured data benefit from real-time linked visual exploration.

Finally, ARGOS simplifies the process of publishing results for interactive exploration by the broader community. Because it requires only a static HTTP server, without any custom backend code or database, it is ideally suited for long-term, low-cost and zero maintenance academic data releases. Dataset providers can create tailored dashboards that guide users through complex data products, enabling sophisticated forms of interactive data storytelling that are increasingly valuable in large collaborations and public-facing research dissemination.

A particularly important mechanism for enabling such dissemination is the catalogue drill-down workflow. Users begin by visually exploring a high-level catalogue (e.g., a set of galaxies or haloes) and then select specific objects of interest. Fine-grained data for these objects are fetched on demand and viewed in dedicated dashboards. This pattern aligns naturally with the workflow of astronomical surveys and simulation campaigns, where researchers routinely transition between catalogue-level overviews and object-level inspection. Implementing this capability without dedicated server-side code is a key advantage of ARGOS, making it a practical platform for scalable, interactive releases of complex data products from cosmological simulations.

In summary, ARGOS directly responds to a significant unmet need: the absence of powerful, accessible visual analytics tools tailored to the demands of cosmological simulations. It complements existing visualisation software for cosmological simulations by focusing on rapid, intuitive, browser-native exploration rather than heavy computation, facilitating hypothesis generation and knowledge discovery. By lowering barriers to adoption, enhancing researchers' ability to investigate complex datasets, and enabling interactive dissemination of results, ARGOS contributes to a broader cultural shift toward embracing exploratory, visual, and user-centred analysis as a valuable component of data-driven sciences.

We consider ARGOS to have reached a level of maturity that makes it ready for adoption by the broader scientific community. In particular, its streamlined and intuitive user experience is expected to lower the barrier to entry for a wide range of users, including students and researchers with limited experience in data visualisation tools. We expect this accessibility to be one of its key strengths. Hence, care must be taken to avoid excessive feature growth that could compromise its simplicity, intuitiveness, and overall user experience.

In the short term, development will focus on expanding the range of supported visualisations and further improving scalability. Additional visualisation types, such as density plots, hexbin maps, and dimensionality-reduction views, would provide users with a richer set of exploratory tools for analysing high-dimensional datasets. Furthermore, browser-side performance can be improved through dynamic subsampling and level-of-detail techniques, enabling interactive exploration of increasingly large datasets while maintaining responsive rendering and interaction.

In the longer term, a promising direction is the introduction of collaborative features, including session sharing, saving and sharing of queries, annotations, and persistent bookmarks. Such capabilities would facilitate collaboration between researchers, improve reproducibility of exploratory analyses, and strengthen ARGOS as a platform for communicating and disseminating scientific results.

\bibliography{references.bib}

\end{document}